\documentclass[%
 reprint,
 amsmath,amssymb,
 aps,
]{revtex4-2}

\usepackage{graphicx}
\usepackage{dcolumn}
\usepackage{bm}
\usepackage{amsmath}
\usepackage[dvipsnames]{xcolor}

\begin{document}

\preprint{APS/123-QED}

\title{Two-mode hyperradiant lasing in a system of two quantum dots embedded in a bimodal photonic crystal cavity}

\author{Lavakumar Addepalli}
\email{d20034@students.iitmandi.ac.in}
\author{P.K. Pathak}
\email{ppathak@iitmandi.ac.in}
\affiliation{
 School of Physical Sciences, Indian Institute of Technology Mandi, Kamand, H.P., 175005, India}

\date{\today}

\begin{abstract}
We propose a scheme for two-mode hyperradiant lasing in a system comprising two incoherently pumped quantum dots (QDs) coupled to a bimodal photonic crystal cavity. To account for the exciton-phonon interactions, we employ a time-convolutionless polaron-transformed master equation. Both resonant and off-resonant couplings of the QDs to the cavity modes are considered, and we analyze the resulting radiance witness as well as the intra- and inter-mode photon correlations. Furthermore, we demonstrate that phonon-induced cooperative two-mode two-photon processes significantly enhance the radiance witness. This enhancement is quantified by evaluating the emission and absorption rates associated with both single-mode and two-mode two-photon transitions. Finally, we compare the radiance witness and emission spectral linewidths for QDs coupled to single-mode and bimodal cavities, revealing that an increase in radiance witness is accompanied by a narrowing of the emission linewidth.
\end{abstract}

\maketitle

\section{Introduction}
Cooperativity among $N \geq 2$ emitters plays a crucial role in enhancing spontaneous emission, a phenomenon known as Dicke superradiance (or superfluorescence) \cite{Dicke1954, Auffeves2011}, where the emitted radiation scales as $\propto N^2$ compared to that of uncorrelated emitters. Coupling emitters to a cavity can induce emitter-emitter correlations, leading to the realization of both subradiance and superradiance \cite{Reitz2022}. Superradiance has been observed in various systems, including atomic ensembles \cite{Junki2018}, Bose-Einstein condensates (BECs) \cite{Keeling2010}, trapped atoms \cite{Reimann2015}, and quantum dots (QDs) coupled to cavities \cite{Scheibner2007, Leymann2015}.

In the strong coupling regime, where the spontaneous emission rate $\gamma$ and the cavity mode decay rate $\kappa$ are smaller than the emitter-cavity coupling strength $g$ ($\gamma, \kappa < g$), the system can exhibit ``Hyperradiance", characterized by emission intensity exceeding $N^2$. This phenomenon has been predicted in systems of coherently driven two-level atoms strongly coupled to a single-mode cavity \cite{Pleinert2017}, and is often accompanied by signatures of non-classicality \cite{Xu2017}. However, average number of photons in cavity mode remains negligible under hyperradiant emission. More recently, we have predicted hyperradiance in two quantum dots when cavity mode has maximum number of photons\cite{Addepalli2024}.

Furthermore, this enhanced radiation emission can lead to linewidth narrowing of the cavity output, which has significant applications in areas such as sensing \cite{Degen2017}, precision metrology \cite{Giovannetti2011}, and atomic clocks \cite{Meiser2009, Norcia2016}.

Moreover, systems with emitters coupled to bimodal cavities exhibit inter-mode correlations that facilitate enhanced emitted radiation. In atomic systems, such interactions with bimodal cavities have been shown to lead to two-photon scattering processes \cite{Richter2025} and pronounced two-atom, two-photon Rabi oscillations \cite{Pathak2004}. Additionally, single-photon subtraction in systems comprising two three-level quantum dots (QDs) coupled to a bimodal waveguide has been shown to enable multi-photon Fock state generation \cite{Pasharavesh2024, Pasharavesh2025}. Further, sub-Poissonian light generation in QD systems coupled to bimodal planar photonic crystal cavities \cite{Majumdar2012} demonstrated, and also photon blockade and enhancement of anti-bunching effects in microdisk resonators where emitters are coupled to whispering gallery modes (WGMs) \cite{Barak2008, Liu2016, Xie2016}. Also, unconventional photon blockade is realized in the system with a QD coupled to two orthogonally polarized micropillar cavity modes \cite{Snijders2018}. Lukin et al. \cite{Lukin2023} reported superradiance and explored multimode interference effects in a system of two silicon carbide color centers weakly coupled to the WGMs of a microdisk resonator. In an earlier study, Verma et al. \cite{Verma2020} investigated phonon-induced cooperative two-mode two-photon emission in a system of two quantum dots coupled to a bimodal photonic crystal cavity. In this work, we show that cooperative two-mode two-photon lasing contribution leads to ``Hyperradiance" in the system where two quantum dots are incoherently pumped and strongly coupled to a bimodal cavity. Such a system is experimentally realizable with site controlled quantum dots coupled to bimodal photonic crystal cavities \cite{Majumdar2012}, bimodal micro-pillar cavities \cite{Snijders2018,Heermeier2022}, WGMs of microdisk resonators \cite{Lukin2023}.

Semiconductor quantum dots (QDs) coupled to photonic crystal cavities form on-chip cavity quantum electrodynamics (cavity-QED) systems that enable the exploration of quantum phenomena. These systems can operate in the strong coupling regime \cite{Hennessy2007}, facilitating coherent light–matter interactions at the single-photon level. This regime allows for the observation of Rabi oscillations \cite{Wilson2002,Nomura2010}, photon blockade \cite{Tang2015, Hou2019}, the generation of single photons \cite{Strauf2011}, and the implementation of quantum gates \cite{Najer2019}. Due to their scalability and compatibility with integrated platforms, these systems hold great promise for the advancement of quantum photonic technologies.

Despite their potential, in semiconductor QD–cavity QED systems, the QD-exciton interaction with phonons is inevitable leading to dephasing. In this regime, the dominant mechanism of exciton–phonon interaction is the coupling of QD excitons to longitudinal acoustic (LA) phonons via the deformation potential, which outweighs both piezoelectric and polar coupling to optical phonons \cite{Krummheuer2002,Ramsay2010}. Exciton–phonon interactions (EPI) lead to both coherent and incoherent phonon-assisted processes, including excitation transfer \cite{Majumdar2012phonon, Calic2017}, cavity mode feeding \cite{Calic2011,Hennessy2007}, enhanced exciton excitation and excitation-induced dephasing \cite{Roy2011,Ulrich2011}, as well as temperature-dependent renormalization of the coupling strength \cite{Ramsay2010,Wei2014}.

In this work, we demonstrate that electron-phonon interaction (EPI) plays a significant role in enhancing the emitted radiation in an off-resonantly coupled system consisting of two quantum dots (QDs) embedded in a bimodal cavity. We investigate this enhancement by analyzing the cavity photon statistics, including the mean photon number, radiance witness, and both inter- and intra-mode zero-time-delay second-order photon correlations. Furthermore, we derive the laser rate equations to evaluate the contributions of single-mode and two-mode two-photon processes to the cavity mode population, and explore the relationship between the radiance witness and the emission linewidth. We also studied the effect of the second mode coupling to QDs, considering both the resonant and off-resonant cases. In Section II, we present the model system and derive the corresponding master equation. The results are discussed in Section III, and conclusions are drawn in Section IV.

\section{Incoherently pumped two QD-bimodal cavity system}

\begin{figure}
    \centering
    \includegraphics[width=0.75\columnwidth]{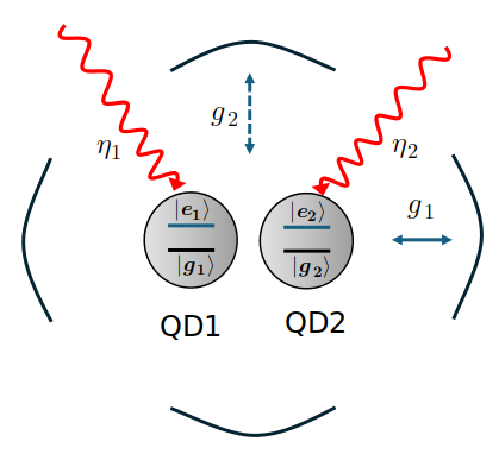}
    \caption{ Schematic figure showing two quantum dots (QDs) coupled to a bimodal cavity with coupling strength, $g_1$ to mode 1, $g_2$ to mode 2 and are incoherently pumped with strengths, $\eta_1$, $\eta_2$.}
    \label{fig:Fig1}
\end{figure}

Here we consider two identical QDs, incoherently driven and coupled to bimodal photonic crystal cavity, where the cavity modes interact via QDs, with no direct coupling between the modes. The Hamiltonian for the system in the rotating frame of QDs transition frequency is given as,

\begin{equation}
    \begin{split}
         H = & -\hbar \delta_1 a_1^\dagger a_1-\hbar \delta_2 a_2^\dagger a_2
         \\&+\hbar \sum_{i=1,2} g_i(\sigma_1^+a_i+ a_i^\dagger \sigma_1^-)+g_i(\sigma_2^+a_i+ a_i^\dagger\sigma_2^-)+ H_{ph},
    \end{split}
\end{equation}

Here, $\delta_i=\omega_D-\omega_{c_i}$ where $\omega_D$ is the QD exciton transition frequency and $\omega_{c_i}$ is the \textit{i}th cavity mode frequency. $a_i$ is the creation operator for \textit{i}th cavity mode and $\sigma_i^-=|g_i\rangle\langle e_i|$, the \textit{i}th QD operator. At cryogenic temperatures, exciton-phonon interactions in these semiconductor cavity QED systems are predominantly via deformation potential coupling. The exciton-phonon interaction Hamiltonian, $H_{ph}=\hbar \Sigma_k (\omega_k a_k^\dagger a_k+\Sigma_i\lambda_k^i\sigma_i^+\sigma_i^-(b_k +b_k^\dagger))$. Here, $\lambda_k^i$ is the coupling strength between the \textit{i}th QD exciton and the \textit{k}th phonon bath mode. $b_k$ is the annihilation operator of the \textit{k}th phonon bath mode. Further, we make a polaron transformation, taking the system to the polaron frame, $\Tilde{H}=e^S H e^{-S}$ where $S=\Sigma_k \Sigma_{i=1,2} \sigma_i^+\sigma_i^- \frac{\lambda_k^i}{\omega_k} (b_k-b_k^\dagger)$. The transformed Hamiltonian, $\Tilde{H}=H_s+H_b+H_{sb}$. Here,

\begin{equation}
    \begin{split}
        H_s =&-\hbar \Delta_{1} a^\dagger a_1 - \hbar \Delta_{2} a_2^\dagger a_2 +\langle B\rangle X_g 
    \end{split}
\end{equation}

\begin{equation}
    H_b = \hbar \Sigma_k \omega_k b_k^\dagger b_k
\end{equation}

\begin{equation}
    H_{sb} = \zeta_g X_g + \zeta_u X_u
\end{equation}

The phonon bath induced polaron shifts, $\Sigma_k \frac{(\lambda_k^i)^2}{\omega_k}$ are absorbed in the detunings, $\Delta_1$ and $\Delta_2$. Throughout our calculation, we assume $\lambda_k^1=\lambda_k^2$, i.e., both the QDs are equally coupled to the \textit{k}th phonon bath mode. The system operators are given by, $X_g= \hbar \Sigma_{i=1,2} (g_i\sigma_1^+a_i + g_i\sigma_2^+a_i)+H.C.$ and $X_u=i\hbar\Sigma_{i=1,2}(g_i\sigma_i^+a_i+g_i\sigma_2^+a_i)+H.C.$ and the bath fluctuation operators are, $\zeta_g=\frac{1}{2}(B_+ + B_- - 2\langle B\rangle)$ and $\zeta_u=\frac{1}{2i}(B_+ - B_-)$. Here, $B_\pm = e^{\pm\Sigma_k \frac{\lambda_k}{\omega_k}(b_k-b_k^\dagger)}$ are the phonon displacement operators. The mean phonon displacement is, $\langle B\rangle = \langle B_+\rangle = \langle B_-\rangle = \exp{[-\frac{1}{2} \int_0^\infty \frac{J(\omega)}{\omega}coth(\beta \hbar\omega/2)]}$ assuming bath is in thermal equilibrium at temperature, T having Bose-Einstein distribution. Here we consider super-ohmic phonon spectral function ,$J(\omega)=\Sigma_k(\lambda_k^{i})^2\delta(\omega-\omega_k)=\alpha_p\omega^3\exp[-\frac{\omega^2}{2\omega_b^2}]$, takes the latter form in continuum limit\cite{Wilson2002,Roy2011}. $\langle B\rangle$ is equal to 0.9, 0.84 and 0.72 for T=5K, T=10K and T=20K respectively.

Further, we derive the master equation for the system by including the residue term after making the polaron transformation, $H_{sb}$ is treated using the Born-Markov approximation \cite{Nazir2016, Mahan1990}. We also phenomenologically incorporated incoherent processes such as cavity decay, QD exciton spontaneous emission, incoherent pumping, and pure dephasing via Lindblad superoperators \cite{Carmichael1999}. The master equation for the QDs-cavity system after tracing over the phonon bath modes is given by,

\begin{equation}
    \begin{split}
    \dot{\rho_s} = &-\frac{i}{\hbar}[H_s,\rho_s]-L_{ph}\rho_s-\Sigma_{j=1,2}\frac{\kappa_j}{2}L[a_j]\rho_s\\&-\Sigma_{i=1,2}(\frac{\gamma_i}{2}L[\sigma_i^-]+\frac{\eta_i}{2}L[\sigma_i^+]+\frac{\gamma_i'}{2}L[\sigma_i^+\sigma_i^-])\rho_s
    \end{split}
    \label{eqn:ME}
\end{equation}

Here, the second term on the right-hand side of Eq. \ref{eqn:ME}, corresponds to exciton phonon-induced processes' effect on system dynamics given by,

\begin{equation}
    \begin{split}
        L_{ph}\rho_s = &\frac{1}{\hbar^2}\int_{0}^{\infty}d\tau \Sigma_{j=g,u}G_j(\tau)\\&\times[X_j(t),X_j(t,\tau)\rho_s(t)]+H.C.
    \end{split}
    \label{eqn:Lph}
\end{equation}

where, $X_j(t,\tau)=e^{iH_s\tau}X_j(t)e^{-iH_s\tau}$. $G_j(\tau)=\langle \zeta_j(0)\zeta_j(\tau)\rangle_{bath}$ are the bath correlation functions $G_g(\tau)=\langle B \rangle^2{\cosh(\phi(\tau)-1)}$, $G_u(\tau)=\langle B \rangle^2\sinh(\phi(\tau))$ 

\section{Results}

\begin{figure}
    \centering
    \includegraphics[width=\columnwidth,height=\columnwidth]{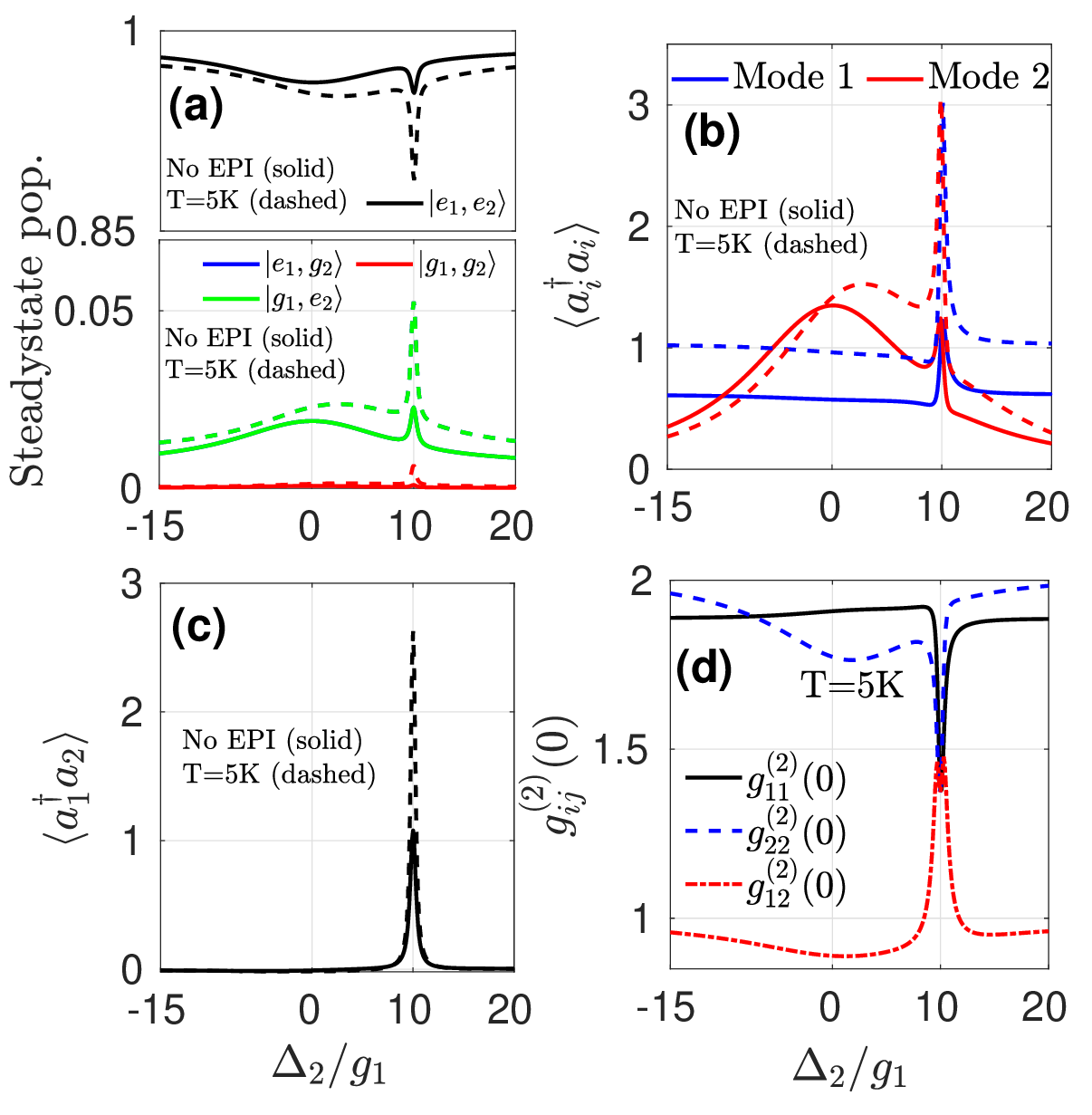}
    \caption{(Color online) Varying second cavity mode detuning w.r.t. QDs, $\Delta_2$ for $\Delta_1=10.0g_1$ and pumping rate, $\eta_1=\eta_2=\eta=25.0g_1$.(a) Steady-state population of collective QD states, $|e_1,e_2\rangle$ (black), $|e_1,g_2\rangle$ (blue), $|g_1,e_2\rangle$ (green) and $|g_1,g_2\rangle$ (red). Under the consideration of identical QDs, the population of $|e_1,g_2\rangle$ and $|g_1,e_2\rangle$ are overlapped. (b) mean photon number of mode 1, $\langle n\rangle_1$ (red) and mode 2, $\langle n\rangle_2$ (blue) (c) correlation between the cavity modes, $\langle a_1^\dagger a_2\rangle$ for ``No EPI" (dashed) and T=5K (solid) (d) intra and inter-mode zero-time delay second order correlation function for T=5K case. Considered system parameters, mode 2 coupling strength, $g_2=1.0g_1$, cavity decay rates, $\kappa_1=\kappa_2=\kappa=0.5g_1$, spontaneous decay rate of the QDs are $\gamma_1=0.01g_1$, $\gamma_2=0.01g_1$ respectively and the pure dephasing rates are $\gamma_1'=\gamma_2'=0.01g_1$.}
    \label{fig:Fig2}
\end{figure}

We consider both the QDs to be equally coupled to the cavity modes within the strong coupling regime, $\gamma_i,\kappa_i<g_i$ with $\gamma_i=0.01g_1$ and $\kappa_1=\kappa_2=0.5g_1$. We study the two cases where the QDs are off-resonantly and resonantly coupled to the cavity modes. In the off-resonant case, the phonon-induced effects play a significant role on the dynamics of the QD state populations and cavity photon statistics. 

\paragraph{Off-resonant case:} In Fig. \ref{fig:Fig2} we consider cavity mode 1 detuning $\Delta_1=10.0g_1$ and compare the results with and without exciton-phonon interaction (No EPI). We show the results for steady-state populations of the collective QD states along with cavity photon statistics for the incoherent pumping rate, $\eta=25.0g_1$, while the detuning of the cavity mode 2, $\Delta_2$ is varied. From Fig. \ref{fig:Fig2}(a), it is evident that when both the modes are equally detuned i.e., $\Delta_2=\Delta_1=10.0g_1$, there is a noticeable transfer of population from the state $|e_1,e_2\rangle$ showing dip to the states $|e_1,g_2\rangle$, $|g_1,e_2\rangle$ and $|g_1,g_2\rangle$ showing sharp peak. This leads to the population of the cavity modes ($\langle n\rangle_i$, $i=1,2$) as shown in the  Fig. \ref{fig:Fig2}(b). The population of the states $|e_1,g_2\rangle$ and $|g_1,e_2\rangle$ states overlap since we consider QDs are identical and their coupling strength to the cavity modes are equal, $g_2=g_1$. The mean photon number in mode 1, $\langle n\rangle_1$ remains nearly constant until the detuning $\Delta_2$ approaches $10.0g_1$, where there is a sharp peak. In contrast, the mean photon number in mode 2, $\langle n\rangle_2$ increases gradually as the cavity mode is tuned to resonance with the QDs, $\Delta_2=0.0$, eventually attaining a broadened peak. This peak shifts slights towards right due to phonon induced effects for T=5K case. At $\Delta_2=10.0g_1$, a peak also appears in the mean photon number of mode 2, $\langle n\rangle_2$ reaching a value equal to that of $\langle n\rangle_1$. Further increase in $\Delta_2$ away from $\Delta_1=10.0g_1$, the mean photon number of the mode 1 decreases sharply and attains a constant value and while, that of mode 2 decreases gradually. In this off-resonant scenario, compared to No EPI case (solid) there is large population transfer and increased mean photon number for T=5K case (dashed). In addition to the enhanced cavity mode population at $\Delta_2=10.0g_1$, Fig. \ref{fig:Fig2}(c) shows the establishment of correlation between both the cavity modes, $\langle a_1^\dagger a_2\rangle$ displaying a sharp peak.

The results for intra- and inter-mode zero-time delay second order photon correlation functions are given in Fig. \ref{fig:Fig2}(d). We can see that for $\Delta_2=\Delta_1=10.0$, there is sharp dip in the intra-mode correlation, $g_{ii}^{(2)}(0)$, here $i=1,2$ driving cavity field from thermal to lasing behavior. We also notice peak in inter-mode correlation function, $g_{12}^{(2)}(0)$ attaining value equal to $g_{ii}^{(2)}(0)$. This implies that each cavity mode is equally correlated with itself and with the other mode when $\Delta_2=\Delta_1$. This correlation contributes to the enhancement of cavity mode photon number and facilitates co-operative two-mode two-photon emission \cite{Verma2020} and is discussed in the latter part of the section. The results show that the influence of the presence of mode 2 on the photon statistics of mode 1 is maximum when both are equally tuned with respect to QDs.

\begin{figure}
    \centering
    \includegraphics[width=\columnwidth]{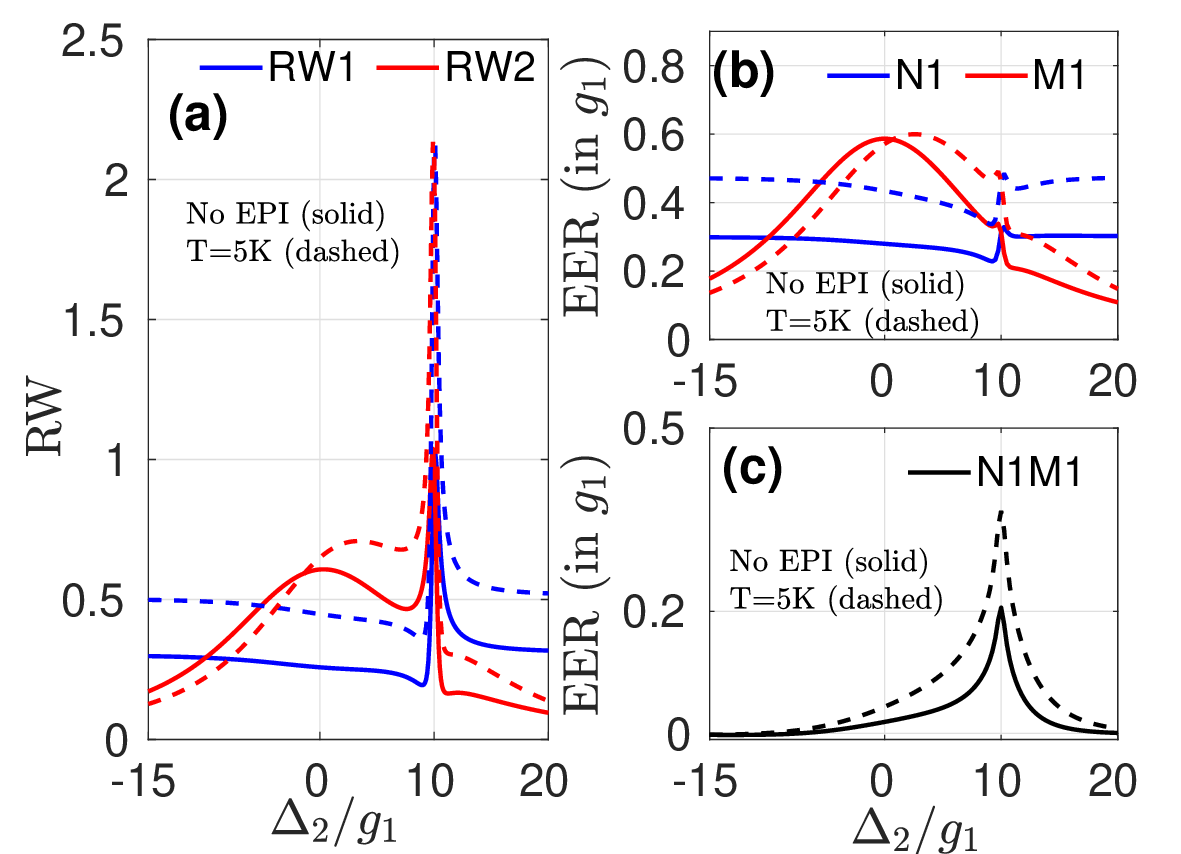}
    \caption{Radiance witness and excess emission rates (EERs) for ``No EPI" (solid) and T=5K (dashed) cases are given.(a) Radiance witness of mode 1, RW1 (blue), mode 2, RW2 (red) for  In off-resonant case, EPI interaction leads to enhanced emission into the cavity modes. (b) Single-photon EER into mode 1 (N1, blue), mode 2 (M1,red) and (c) two-mode two-photon EER (N1M1, black). The parameters are same as in Fig.\ref{fig:Fig2}.}
    \label{fig:Fig3}
\end{figure}

In Fig. \ref{fig:Fig3}, we present the results for the radiance witness and single, two-mode two-photon excess emission rates  varying cavity detuning, $\Delta_2$ and compared the results for No EPI (solid) and temperature, T=5K (dashed) cases. Fig. \ref{fig:Fig3} (a) presents the result for the radiance witness (RW) defined as, RW$=\frac{\langle n\rangle_2-2\times\langle n\rangle_1}{2\times\langle n\rangle_1}$, where $\langle n\rangle_2$ denotes the mean photon number when two emitters are coupled to the cavity mode, and $\langle n\rangle_1$ corresponds to that for a single emitter \cite{Pleinert2017}. The value of RW characterizes the nature of collective emission: RW$<0$ indicates subradiance, RW$>0$ enhanced emission, RW$=1$ superradiance and RW$>1$ corresponds to Hyperradiance. 

Furthermore, we note that when both the modes are equally detuned, $\Delta_2=\Delta_1=10.0g_1$, the system exhibits Hyperradiant behavior with cavity modes are predominantly populated through single photon emission and cooperative two-mode two-photon emission processes. Fig. \ref{fig:Fig3} (a) shows the results of the radiance witness (RW) for both the cavity modes using the same parameters as in Fig. \ref{fig:Fig2}. Similar to the behavior observed in the mean photon number of mode 1, Fig. \ref{fig:Fig2} (b), the radiance witness of the mode 1 (RW1) also exhibits a peak at the mode 2 detuning, $\Delta_2=10.0g_1$. The radiance witness of mode 2, RW2 shows a gradual increase in its value with a broadened peak around $\Delta_2\approx 0$ and attains sharp peak at $\Delta_2=10.0g_1$. In this off-resonant scheme, phonon-induced effects enhance the radiance witness values, resulting in Hyperradiance with RW$\approx 2.1$ compared to the ``No EPI" case where RW$\approx 1.0$, indicating Superradiance. This enhancement in RW is a result of cooperative two-mode two-photon emission into the cavity modes. In the following, we analyze the individual contribution of single-photon and two-mode two-photon processes to the cavity mode population by evaluating their emission and absorption rates. We plot the results for the difference between the emission and absorption rates, referred to as the "Excess Emission Rate (EER)".

We have computed the single and multi-photon absorption and emission rates contributing to the cavity mode mean photon number, as given in Fig. \ref{fig:Fig3} (b) and (c) for T=5K (dashed), ``No EPI" (solid) cases. We follow the standard procedure of Scully-Lamb quantum theory of lasing to evaluate the emission and absorption rates. Specifically, we employ the simplified master equation (SME) of the system, detailed in Appendix \ref{sec:AppendixA} to determine the rate equations of the density matrix elements, $\langle i_1,j_2,m,n|\rho_s|i_1,j_2,p,q\rangle$ where $i, j=e,g $ denote QD states and $m, p$ ($n, q$) represent the photon numbers in mode 1 (mode 2). We then trace over the QD states to obtain the rate equation for the cavity fields \cite{Addepalli2024, Hazra2024}.

\begin{widetext}
    \begin{equation}
    \begin{split}
        \dot{P}_{n,m} = &-\alpha_{n,m} P_{n,m}+G^{11}_{n-1,m-1}P_{n-1,m-1} 
        +G^{10}_{n-1,m}P_{n-1,m}+G^{01}_{n,m-1}P_{n,m-1}
        +A^{11}_{n+1,m+1}P_{n+1,m+1}+A^{10}_{n+1,m}P_{n+1,m}
        \\&+A^{01}_{n,m+1}P_{n,m+1}
        +\kappa_{1}(n+1)P_{n+1,m}- \kappa_{1}nP_{n,m}
        +\kappa_{2}(m+1)P_{n,m+1} - \kappa_{2}mP_{n,m}.
    \end{split}
    \label{eqn:cavityRateEqn}
\end{equation}
\end{widetext}

The probability of having $n$, $m$ photons in 1st, 2nd cavity modes respectively is given by, $P_{nm}=\Sigma_i \langle i,n,m|\rho_s|i,n,m\rangle$, while  $\alpha_{n,m}=\Sigma_i \alpha_{i,n,m}\langle i,n,m|\rho_s|i,n,m\rangle$. The emission and absorption coefficients are given by,, $G_{n,m}^{ab}P_{n,m}=\Sigma_i G_{i,n,m}^{ab}\langle i,n,m|\rho_s|i,n,m\rangle$ and $A_{n,m}^{ab}P_{n,m}=\Sigma_i A_{i,n,m}^{ab} \langle i,n,m|\rho_s|i,n,m\rangle$ where $i={x,y,g}$. The coefficients $\alpha_{i,n,m}$, $G_{i,n,m}^{ab}$ and $A_{i,n,m}^{ab}$ are obtained numerically. The single-photon emission(absorption) rate for the first and second modes are given by, $\Sigma_{n,m} G_{n,m}^{10} P_{n,m}$($\Sigma_{n,m}A_{n,m}^{10}P_{n,m}$) and $\Sigma_{n,m} G_{n,m}^{01} P_{n,m}$($\Sigma_{n,m}A_{n,m}^{01}P_{n,m}$) respectively and two-mode two-photon emission(absorption) rate is given by $\Sigma_{n,m} G_{n,m}^{11} P_{n,m}$($\Sigma_{n,m}A_{n,m}^{11}P_{n,m}$). Subsequently, we define single-photon and two mode two-photon excess emission rates (EER) as the difference between the corresponding emission and the absorption rates. The sign of EER $>0$ or $<0$ represents net emission or absorption occurring in the cavity mode. The results for EERs are presented in Fig. \ref{fig:Fig3} (b) and (c). From Fig.\ref{fig:Fig3} (b), we observe a broadened peak in single photon excess emission rate into mode 2 (M1 curve) when $\Delta_2\approx0.0g_1$ and for $\Delta_2\neq \Delta_1$, both the modes are predominantly populated by single photon processes (N1, M1). However, as the detuning of the mode 2, $\Delta_2$ varied, and when both modes are equally tuned w.r.t QDs i.e., $\Delta_2=\Delta_1=10.0g_1$, the single photon EERs (N1, M1) cross each other. At $\Delta_2=\Delta_1$, both the cavity modes are not only equally populated by single photon processes (N1, M1) but also show significant contribution from cooperative two-mode two-photon process (N1M1) as shown in Fig. \ref{fig:Fig3} (c) as we cam see a sharp peak in ``N1M1" curve. This cooperative two-mode two-photon emission process is responsible for the generation of correlation between both the cavity modes, Fig. \ref{fig:Fig2}(c). We further note that, in comparison to the ``No EPI" case (solid), the two-mode two-photon excess emission rates is twice for T=5K case (dashed) and is attributed to the exciton-phonon interaction in this off-resonantly coupled system.

\begin{figure}
    \centering
    \includegraphics[width=\columnwidth]{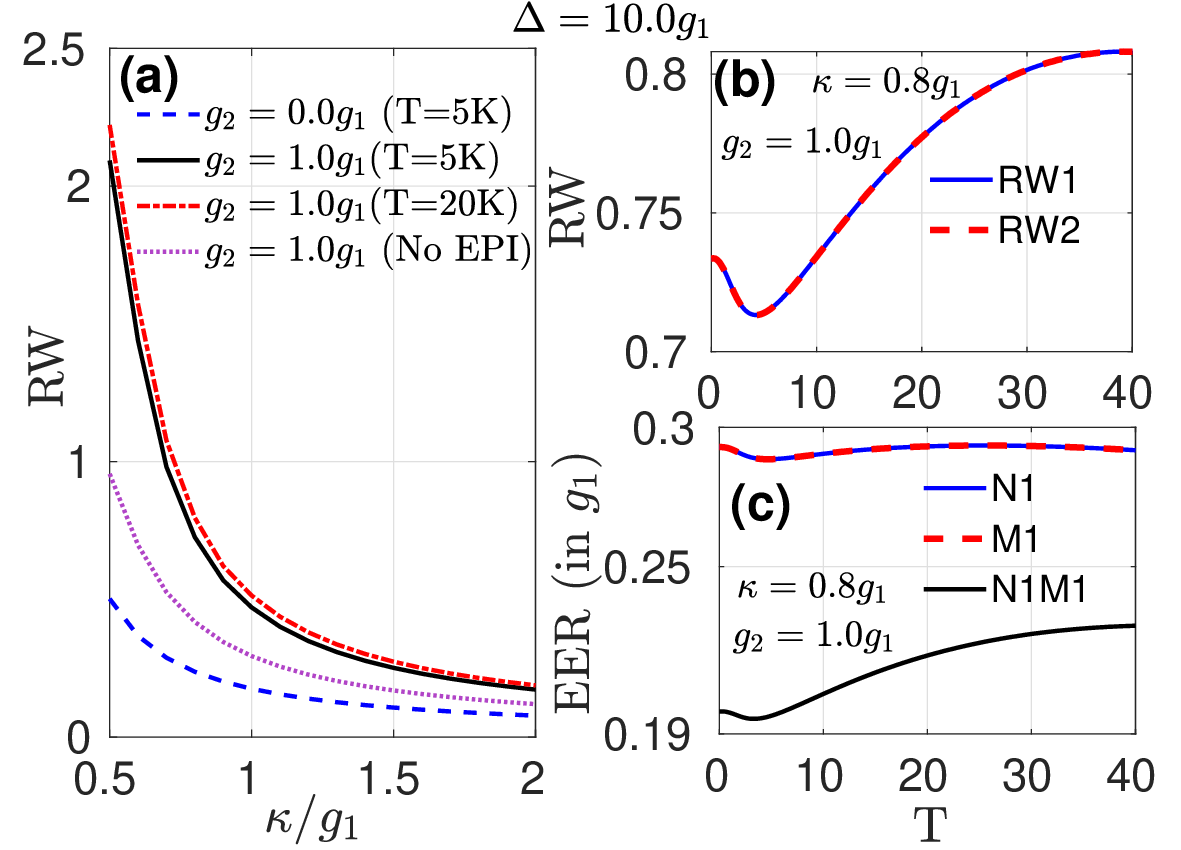}
    \caption{(a) Radiance witness (RW1=RW2=RW) variation with cavity decay rate, $\kappa_1=\kappa_2=\kappa$ is varied. Comparison is made for (i) $g_2=0.0g_1$, single mode T=5K (dashed blue), (ii) $g_2=1.0g_1$, bimodal, T=5K (solid black), (iii) $g_2=1.0g_1$, bimodal T=20K (dash-dotted red) and (iv) $g_2=1.0g_1$, bimodal ``No EPI" (dotted violet) scenarios. In subplot (b), radiance witnesses, RW1 and RW2, and (c) Excess emisssion rates (EERs) are given by varying temperature, T. The cavity mode detunings are fixed at $\Delta_1=\Delta_2=\Delta=10.0g_1$, and cavity decay rates, $\kappa_1=\kappa_2=\kappa=0.8g_1$ considered in subplot (b). The other parameters are same as in Fig. \ref{fig:Fig2}.}
    \label{fig:Fig4}
\end{figure}

Fig. \ref{fig:Fig4} (a) shows the effect of cavity decay rate on the RW for (i) single-mode coupling, $g_2=0.0g_1$, for T=5K, bimodal coupling, $g_2=1.0g_1$ for (ii) T=5K, (iii) T=20K and (iv) ``No EPI" scenarios. With the increase in quality of the cavity modes i.e., decrease in $\kappa$, the bimodal cavity system exhibits ``Hyperradiance" (RW$>1$) when compared to the system with QDs coupled to single-mode for decay rates, $\kappa \lessapprox 0.7g_1$ for T=5K. The radiance witness increases slightly for T=20K and is attributed to the increased two-mode two-photon EER as given in subplot (c). We can see that the RW for ``No EPI" case is far below the T=5K and T=20K cases as the phonon-induced effects dominate.
In Fig. \ref{fig:Fig4} (b) and (c) we show the results for the variation of RW1 and RW2 with temperature, T  by considering both the cavity modes are equally tuned w.r.t QDs i.e., $\Delta=10.0g_1$ and $\kappa=0.8g_1$. We have considered maximum temperature, T upto 40K and for the considered parameters, the system is within the validity regime of polaron theory \cite{Nazir2016}. As the temperature increases, the RW show slight dip initially and rises, which can be attributed to the buildup of correlation between the modes and enhanced contribution of two-mode two-photon emission to the cavity mean photon number. This feature is reflected in the EER results given in Fig. \ref{fig:Fig4} (c) as a function of temperature. As the temperature increases, both single and two-mode two-photon excess emission rates initially show a slight dip. With further increase in T, the single photon EERs show very little variation but the two-mode two-photon EER increases appreciably. Hence, the phonon-assisted two-mode two-photon process strengthens the correlation between the modes, and resulting in the increase of their radiance witnesses, driving the system from superradiance to Hyperradiance.

\begin{figure}
    \centering
    \includegraphics[width=\columnwidth]{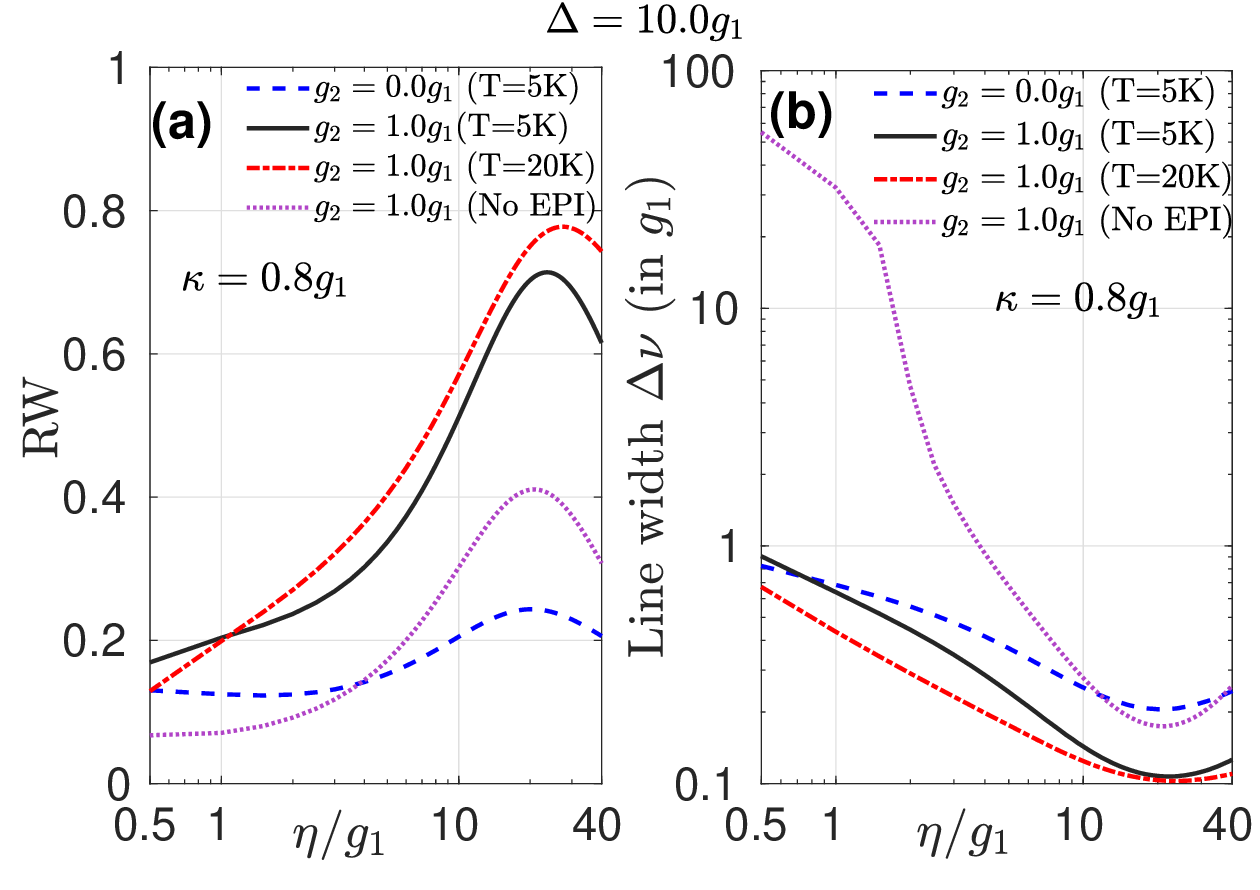}
    \caption{The variation of radiance witness in subplot (a), and linewidth in subplot (b) with increasing incoherent pumping rate, $\eta$. Comparison is made for the scenarios, (i) $g_2=0.0g_1$ (single mode) T=5K, (ii) $g_2=1.0g_1$ (bimodal) T=5K, (iii) $g_2=1.0g_1$ (bimodal), T=20K and (iv) $g_2=1.0g_1$ (bimodal),``No EPI". Color scheme is same as in Fig. \ref{fig:Fig4} (a). The other parameters are same as in Fig. \ref{fig:Fig4} (b).}
    \label{fig:Fig5}
\end{figure}

In Fig. \ref{fig:Fig5}, we show the results of radiance witness and linewidth of the cavity mode emission spectrum with increasing incoherent pumping rate, $\eta$, for equal detuning of the cavity modes $\Delta_1=\Delta_2=\Delta=10.0g_1$ and $\kappa=0.8g_1$. In subplot Fig. \ref{fig:Fig5} (a) we made comparison of the four cases as we did in Fig. \ref{fig:Fig4} (a). The results clearly show that the radiance witness is notably higher for the case of QDs coupled to bimodal cavity compared to the coupling with single mode cavity, indicating enhanced cooperative emission. Futher, the peak value of the RW for ``No EPI" case is almost half that of T=20K case. Additionally, in this off-resonant coupling scheme, the dominant phonon induced processes contribute to increased RW with rise in temperature, T, as shown for T=20K (red dash-dotted line). This enhancement in RW leads to the suppression of the linewidth as given in Fig. \ref{fig:Fig5} (b) as in superradiant lasers \cite{Meiser2009, Kristensen2023}. 

The emission spectrum is obtained using the quantum regression theorem applied to the two-time correlation function, $\langle a^\dagger(t) a(0)\rangle$ \cite{SMTan1999} and the linewidth is evaluated by fitting the spectrum with a Lorentzian function. We can see that for T=5K, as the radiance witness increases and attains peak value $\approx 0.7$, the corresponding linewidth decreases. In particular, for T=20K case, the emission linewidth $\Delta\nu$ reduces to approximately 0.1, which is nearly $85\%$ narrower than the bare cavity linewidth, $\kappa = 0.8g_1$. The results clearly demonstrate that the QDs coupled to bimodal cavities outperform both the single-mode coupling case and bimodal coupling``No EPI" case.

\begin{figure}
    \centering
    \includegraphics[width=\columnwidth]{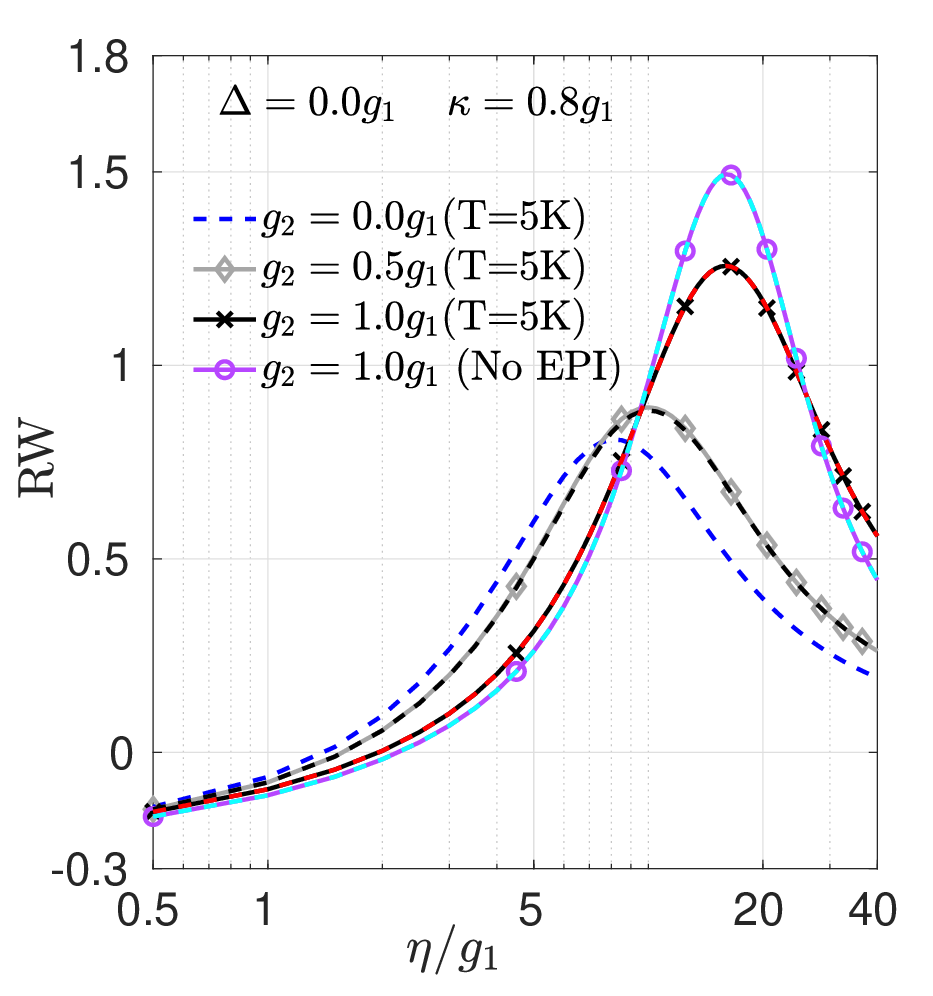}
    \caption{Radiance witness for resonant case, $\Delta_1=\Delta_2=\Delta=0.0g_1$ by varying incoherent pumping rate, $\eta$. Comparison is made between the cases: T=5K  (i) $g_2=0.0g_1$ (Single mode, dashed blue) (ii) $g_2=0.5g_1$ ($\diamond$), (iii) $g_2=0.2g_1$ ($\times$) and (iv) $g_2=1.0g_1$ ``No EPI" ($\circ$). Cavity decay rates, $\kappa=0.8g_1$ and the other parameters are same as in Fig. \ref{fig:Fig2}.}
    \label{fig:Fig6}
\end{figure}

\paragraph{Resonant case:}Fig. \ref{fig:Fig6} shows results for the radiance witness of mode 1 (RW1) and mode 2 (RW2), for the case of QDs resonantly coupled to the cavity modes i.e., $\Delta_1=\Delta_2=0.0g_1$. We vary the incoherent pumping rate, $\eta$, considering both the cavity modes have equal quality factors, $\kappa_2=\kappa_1=0.8g_1$. The results are given for different second mode coupling strengths, (i) $g_2=0.0g_1$ (solid brown) corresponds to single mode scenario, (ii) $g_2=0.5g_1$ (`$\diamond$') at T=5K, (iii) $g_2=1.0g_1$ (`$\times$') at T=5K and (iv) $g_2=1.0g_1$, ``No EPI" (`$\circ$') case. The radiance witness for both the modes overlap even when the second mode is coupling strength, $g_2=0.5g_1$. We note that this is due to the presence of strong correlation between the modes when they are equally detuned, $\Delta_1=\Delta_2$ and having same cavity decay rates, $\kappa_1=\kappa_2$. Increase in the incoherent pumping rate drives the system from subradiant to Superradiant or Hyperradiant regime. For the considered parameters, the peak value of RW1 increases in the presence of the second mode and exceeds unity (RW1$>1$) indicating ``Hyperradiance" over a range of incoherent pumping strengths, $10g_1 \leq\eta\leq 25.0g_1$ as compared to the single mode case whose peak value is RW$\approx 0.8$ (below Superradiant phase). Furthermore, by comparing the results for (ii) and (iii) cases, we can say that with increase in the mode 2 coupling strength, $g_2$, the peak value of RW increases. For resonant case, the phonon induced decoherence leads to the suppression in mean photon number and the peak value of the radiance witness as shown for the ``No EPI" (`$\circ$') and T=5K (`$\times$') cases unlike the off-resonant case where the phonon induced effects enhance the radiance witness, cf., Fig. \ref{fig:Fig3} (a), Fig. \ref{fig:Fig4} (a) and Fig. \ref{fig:Fig5} (a).

\section{Conclusions}
In conclusion, we have demonstrated that when both cavity modes are equally detuned from the quantum dot (QD) transition frequencies, there is a pronounced enhancement in inter-mode correlations and the cooperative two-mode two-photon emission rate, leading to two-mode hyperradiant lasing. In the case of off-resonant coupling, exciton-phonon interactions (EPI) play a crucial role in enhancing the emission rate, as confirmed by comparisons with scenarios where EPI is neglected. Moreover, we have shown that bimodal coupling, in contrast to the single-mode configuration, significantly enhances the radiance witness and supports the emergence of hyperradiant lasing. This enhancement is also associated with a suppression of the laser linewidths. Finally, we examined the resonant coupling regime where exciton-phonon interactions are negligible. In this regime, cavity-mediated two-mode two-photon emission continues to enable hyperradiant lasing. However, in the presence of exciton-phonon interactions at T = 5\,\text{K}, the peak value of the radiance witness is slightly reduced compared to the "No EPI" case, attributed to phonon-induced decoherence.

\begin{widetext}

\appendix
\section{\label{sec:AppendixA} Simplified master equation}

The simplified master equation for the system is given below after making the approximations, $\Delta_1,\Delta_2>>g_i$, and is used to calculate the density matrix element rate equations, thereby calculating single and multi-photon emission and absorption rates.

\begin{equation}
    \begin{split}
        \dot{\rho_s}=& -\frac{i}{\hbar}[H_{eff},\rho_s]-\sum_{i=1}^2\Big(\frac{\kappa_i}{2}L[a_i] -\frac{\gamma_i}{2}L[\sigma_i^-]-\frac{\eta_i}{2}L[\sigma_i^+]-\frac{\gamma_i'}{2}L[\sigma_i^+\sigma_i^-]\Big)\rho_s\\&-\sum_{i,j,k,l=1,i\neq j}^2 \frac{\Gamma_{kl}^{--}}{2}(a_l^\dagger \sigma_j^- a_k^\dagger\sigma_i^-\rho_s-2a_k^\dagger\sigma_i^-\rho_s a_l^\dagger\sigma_j^-+\rho_s a_l^\dagger \sigma_j^- a_k^\dagger \sigma_i^-)+\frac{\Gamma_{kl}^{++}}{2}(\sigma_j^+ a_l \sigma_i^+a_k\rho_s - 2\sigma_i^+a_k\rho_s\sigma_j^+a_l+\rho_s\sigma_j^+a_l\sigma_i^+a_k) \\& -\sum_{i,j,k,l=1}^2 \frac{\Gamma_{kl}^-}{2}(a_l^\dagger\sigma_j^-\sigma_i^+a_k\rho_s - 2\sigma_i^+a_k\rho_s a_l^\dagger\sigma_j^- + \rho_s a_l^\dagger \sigma_j^-\sigma_i^+ a_k)+\frac{\Gamma_{kl}^+}{2}(\sigma_j^+ a_l a_k^\dagger\sigma_i^-\rho_s - 2 a_k^\dagger\sigma_i^-\rho_s\sigma_j^+ a_l + \rho_s\sigma_j^+ a_l a_k^\dagger \sigma_i^-)
    \label{eqn:SME}
    \end{split}
\end{equation}

Where the effective Hamiltonian is,

\begin{equation}
    H_{eff}=H_s-i\hbar\sum_{i,j,k,l=1}^2 \Omega_{kl}^- a_l^\dagger \sigma_j^-\sigma_i^+a_k + \Omega_{kl}^+\sigma_j^+a_l a_k^\dagger\sigma_i^- + \sum_{i,j,k,l=1, i\neq j}^2 i\hbar \Omega_{kl}^{--}a_l^\dagger\sigma_j^-a_k^\dagger\sigma_i^- + H.C.
\end{equation}

The phonon-induced scattering rates are given by,

\begin{equation}
    \Omega_{kl}^{\pm}=\frac{g_k g_l}{2}\int_0^\infty d\tau (G_+e^{\pm i\Delta_k\tau}-G_-^* e^{\mp i\Delta_l\tau})
\end{equation}

\begin{equation}
    \Omega_{kl}^{--}=\frac{g_k g_l}{2}\int_0^\infty d\tau(G_- e^{i\Delta\tau}-G_-^* e^{i\Delta_i\tau})
\end{equation}

\begin{equation}
    \Gamma_{kl}^{\pm}=g_k g_l\int_0^\infty d\tau(G_+ e^{\pm i\Delta_k\tau} + G_+^* e^{\mp i\Delta_l\tau})
\end{equation}

\begin{equation}
    \Gamma_{kl}^{--/++}= g_k g_l\int_0^\infty d\tau (G_- e^{\pm i\Delta_k\tau}+G_-^* e^{\pm i\Delta_l\tau})
\end{equation}

\end{widetext}

\bibliography{TwoQDtwoMode}

\end{document}